# The Analog Front-end Prototype Electronics Designed for LHAASO WCDA

MA Cong(马聪)[1,2]    ZHAO Lei(赵雷) [1,2;1)]

GUO Yu-Xiang(郭宇翔) [1,2] LIU Jian-Feng(刘建峰) [1,2] LIU Shu-Bin(刘树彬) [1,2] AN Qi(安琪) [1,2]

1 State Key Laboratory of Particle Detection and Electronics, University of Science and Technology of China, Hefei, 230026, China,

2 Department of Modern Physics, University of Science and Technology of China, Hefei, 230026, China

**Abstract:** In the readout electronics of the Water Cerenkov Detector Array (WCDA) in the Large High Altitude Air Shower Observatory (LHAASO) experiment, both high-resolution charge and time measurement are required over a dynamic range from 1 photoelectron (P.E.) to 4000 P.E. The Analog Front-end (AFE) circuit is one of the crucial parts in the whole readout electronics. We designed and optimized a prototype of the AFE through parameter calculation and circuit simulation, and conducted initial electronics tests on this prototype to evaluate its performance. Test results indicate that the charge resolution is better than 1% @ 4000 P.E. and remains better than 10% @ 1 P.E., and the time resolution is better than 0.5 ns RMS, which is better than application requirement.

## 1. Introduction

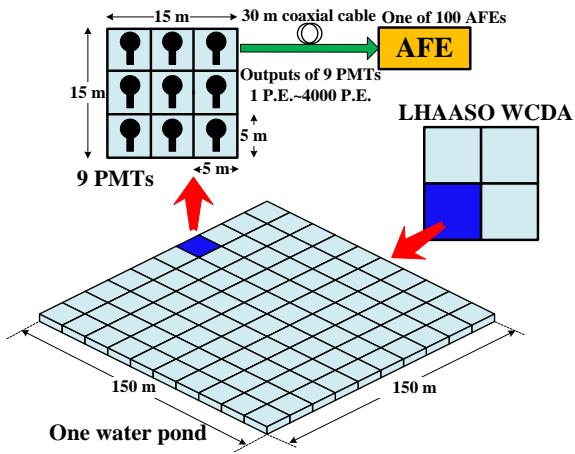

Fig. 1 Structure of one WCDA pond in LHAASO.

The Large High Altitude Air Shower Observatory (LHAASO) aims to detect the gamma ray sources above 30 TeV for the finding of galactic cosmic ray sources to be built at a high altitude in China. LHAASO consists of five kinds of detectors, which are used for different detection energy regions and detection purposes [1, 2]. The Water Cerenkov Detector Array (WCDA) is an important component of LHAASO, and it is an all-sky extensive air shower (EAS) detector array which is sensitive to gamma ray showers above a few hundred GeV [3, 4]. It covers an area of 90,000 m² configured as four individual square water ponds of 150 m×l50 m, marked as the four squares in Fig. 1. In each water pond, 900 photomultiplier tubes (PMTs) are placed under water to detect Cerenkov light emitted by EAS in the water for the event reconstruction in the physics experiment. An analog front-end module above the ponds receives the signals from 9 PMTs within a 15 m × 15 m square through 30-meter cables. The dynamic range of the measurement varies from 1 Photoelectron (P.E.) to 4000 P.E., and both high-resolution charge and time measurement are required; the requirements are listed in Table. 1 [5].

Table. 1. Measurement requirements of the LHAASO WCDA AFE.

| Item | Requirement |
| --- | --- |
| Channel number | 3600 |
| Resolution of charge measurement | 30%@1 P.E. |
| | 3%@4000 P.E. |
| Bin size of time measurement | <1 ns |
| RMS of time measurement | <0.5 ns |
| Dynamic range of time measurement | 2 µs |

*Supported by Knowledge Innovation Program of the Chinese Academy of Sciences (KJCX2-YW-N27), National Natural Science Foundation of China (11175174) and the CAS Center for Excellence in Particle Physics (CCEPP).

1) Email: zlei@ustc.edu.cn

As for charge measurement, there are several methods applied in the readout electronics of physics experiments, such as the digital peak detection method used in the front-end electronics (FEE) of Daya Bay Reactor Neutrino Experiment [6], the time-over-threshold (TOT) principle adopted in the BES III TOF readout electronics [7], and the waveform digitization based on the Switched Capacitor Array (SCA) ASIC like ATWD used in the IceCube experiment [8], etc. To achieve a high charge measurement resolution within a large dynamic range, the charge measurement circuits in our project are designed based on the analog signal shaping, A/D conversion and digital peak detection implemented in the Field Programmable Gate Array (FPGA) devices.

As for time measurement, leading-edge discriminators are combined with Time-to-Digital Converters (TDCs) based on the multi-phase clock interpolation technique in the FPGA to reduce the system complexity [9]. The time walk can be easily corrected in the FPGA using the charge measurement results.

This paper is organized as follows: Section 2 describes the whole system architecture. Section 3 and 4 narrate the kernel circuits design, as well as the simulation based on PSpice of the charge and time measurement circuit, respectively; Section 5 presents a series of electronics test results in the laboratory; in Section 6, we conclude the paper and summarize what has been achieved.

## 2. Architecture of the Analog Front-end electronics

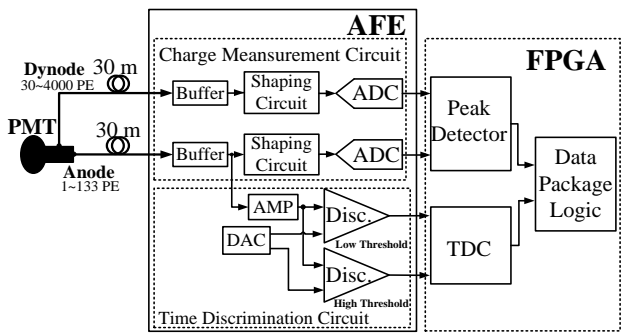

Fig. 2 Structure of the AFE.

According to the aforesaid measurement requirements, the main design considerations of the AFE include:

1) Large dynamic range measurement. As mentioned above, the input signal ranges from 1 P.E. to 4000 P.E.; it is a great challenge to achieve a high measurement resolution and good linearity over the whole range. To address this issue, we designed two readout channels for each PMT, which covers different measurement ranges. One channel is for the Anode, and the other is for a Dynode, as shown in Fig. 2. These two channels have a similar circuit structure, and achieve a dynamic range of 1 ~ 133 (1 P.E. ~ 133 P.E. for the Anode; 30 P.E. ~ 4000 P.E. for the Dynode), with which the range of 1~4000 is successfully covered. Considering that the slew rate can be guaranteed for both the small and large signals in the Anode channel, time measurement over the whole dynamic range can be achieved with the Anode channel alone. However, we should also consider that the noise or interference in the baseline of the large input signal would cause deterioration of time measurement resolution at the threshold of 1/4 P.E. Therefore, two discriminators are designed with a low (1/4 P.E.) and a high threshold (5 P.E. in the test, user controlled).

2) High-precision impedance matching requirement. Since the dynamic range is from 1 P.E. to 4000 P.E. and the cable connecting the PMT to the subsequent AFE is as long as 30 meters, reflection of the large signal would introduce interference on the measurement of small signals. Therefore, good impedance matching has to be considered.

3) High-resolution charge and time measurement of small signals. The amplitude of the 1 P.E. signal from the Anode is as low as 3 mV on the 50 $\Omega$ resistor, and is further attenuated during propagation through the 30 m cable that also reduces the slope at threshold-level crossing of the signal. Therefore, how to achieve good signal noise ratio (SNR) is a great concern.

The structure of the AFE is shown in Fig. 2. The following sections 3 and 4 will introduce the design of the charge and time measurement circuits of the AFE, with circuit analysis, calculation, as well as simulation.

# 3 The design of the charge measurement circuit

## 3.1 The circuit structure and SNR analysis

In the design of high-resolution charge measurement circuit, the charge sensitive amplifier (CSA) is a common choice as the pre-amplifier, which features very low noise configuration. The CSA is often followed by a shaping circuit consisting of a CR-RC$^m$ filter to form a quasi-Gaussian signal, and an excellent SNR can be achieved with the best time constant selected [10].

However, it is low equivalent impedance at the input of the CSA, which varies with input signal frequency (~ $1/j\omega C$). Therefore, the CSA is unsuitable as the pre-amplifier in our design which requires good impedance matching. For this reason, we use a high-precision 50 $\Omega$ resistor to achieve impedance matching, marked as $R_t$ in Fig. 3. A current signal from PMT is converted to a voltage signal which is amplified by the pre-amplifier ($A_1$ in Fig. 3), and further shaped to a quasi-Gaussian signal by the following RC$^n$ filter. Considering the high input impedance of $A_1$, a good impedance matching precision can be achieved. Furthermore, for higher circuit stability, we adopt the one-pole operation for each stage of the RC$^n$ filter rather than multi-pole configuration, such as the Sallen-key structure [11].

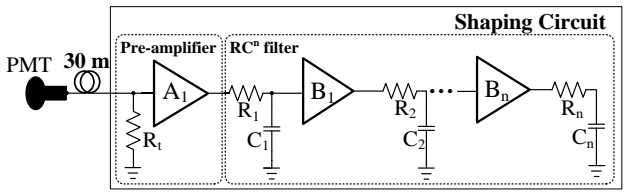

Fig. 3. Structure of the shaping circuit.

To optimize the charge measurement resolution, we need to carefully consider the value of n (the order of the RC$^n$ filter) and $\tau$ (the time constant of the RC$^n$ filter). As for charge measurement, the greatest challenge is to achieve a high measurement resolution of the smallest signal (1 P.E.), so we will deliberate on the noise analysis of the Anode channel which covers the signal ranging from 1 P.E. to 133 P.E.

Generally, the input noise spectral density of the operational amplifier (OPA) consists of two regions: one is dominated by the flicker noise, and the other is dominated by the thermal noise (white noise). The shared border of these two regions is defined as "corner frequency" (usually 1 Hz ~ kHz) [12, 13]. Compared with the passband of the RC$^n$ filter (~ MHz) in this design, the corner frequency is extremely low, which means that the input noise of the pre-amplifier can be considered as white noise (the input noise power spectral density is marked as $a^2$ in the following discussion). Since the noise of the pre-amplifier contributes majority of the circuit noise, to simplify the analysis we that would be caused by other OPAs in Fig. 3 ($B_1$~$B_n$).

The response of a PMT to a light pulse can be considered as its characteristic impulse response if the light pulse duration is negligible compared to this response [14], and it is just the case in the signal dynamic range of the Anode channel. By using the transfer function of the system and inverse Laplace transform, the impulse response of the RC$^n$ filter in time domain and the peak time ($t_M$) can be calculated as:

$$V_o(t) = \frac{A_v}{\tau^n} \frac{1}{(n-1)!} t^{n-1} e^{-\frac{t}{\tau}} u(t) \tag{1}$$

$$t_M = (n-1)\tau, \tag{2}$$

where $A_v$ is the loop gain of $A_1$.

Obviously, $t_M$ for RC$^1$ filter is zero, rendering it useless in the A/D conversion and digital detection method. By integrating the output noise power spectral density over frequency, the output noise voltage (RMS) can be calculated. Meanwhile, the peak value of output signal can be obtained according to Eq. (1). And then, the SNR of the RC$^2$, RC$^3$ and RC$^4$ filter can be separately calculated, as in:

$$\eta(RC^2) = \frac{QR_t}{ea} \sqrt{\frac{4}{\pi\tau}} \tag{3}$$

$$\eta(RC^3) = \frac{2QR_t}{e^2a} \sqrt{\frac{16}{3\pi\tau}} \tag{4}$$

$$\eta(RC^4) = \frac{27QR_t}{6e^3a} \sqrt{\frac{32}{5\pi\tau}}, \tag{5}$$

where $Q$ is the charge of the input signal.

With the above equations, we can plot the SNR (normalized according to the SNR of the $RC^2$ filter) with different value of $\tau$ for these 3 filters, as shown in Fig. 4. There is an obvious negative correlation between SNR and $\tau$, and the $RC^2$ filter has the best SNR performance. Similarly, it can also be proven that the higher the filter order is, the worse the SNR is.

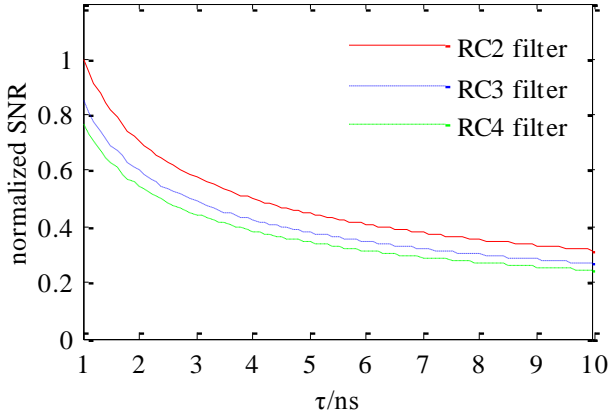

Fig. 4. The SNR calculation results of the $RC^n$ filter (n=2, 3, 4).

To verify the above analysis, PSpice simulation was conducted. We downloaded the Pspice models of AD8000 and AD8058 from the ADI's website, which are used as $A_1$ and other OPAs in the filter ($B_1 \sim B_n$), and then conducted a series of simulations. The input current signal is generated according to the smallest PMT output signal (1 P.E., rise time: ~4 ns, fall time: ~12 ns). As Fig. 5 shows, the simulation results concord well with the above analysis results, which also indicate that the best SNR can be obtained with the $RC^2$ filter.

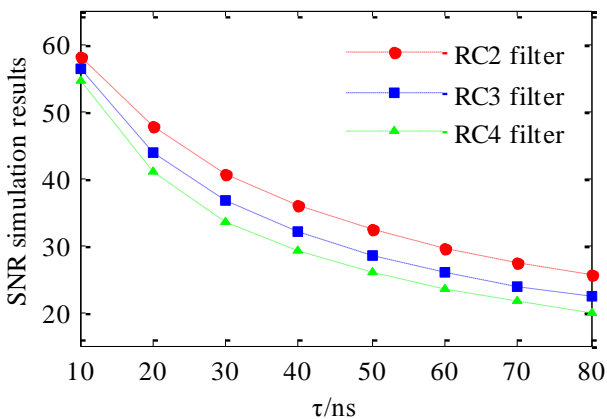

Fig. 5 The SNR PSpice simulation results of the $RC^n$ filter for the 1 P.E. input PMT signal (n=2, 3, 4).

The analysis and simulation results also indicate that $\tau$ should be set smaller to acquire better SNR performance. However, according to Eq. (1), $\tau$ also determines the FWHM of the filter output signal, which will affect the A/D conversion and digital peak detection procedure. It will be discussed in the following sub-section.

### 3.2 peak error

The quasi-Gaussian signal of the $RC^2$ filter output is digitalized by the following Analog-to-Digital Converter (ADC), the output of which are fed into the FPGA for the peak detection, i.e. the maximum value ($V_p$) of the digitized data stream out of the ADC. Since the sampling clock of the ADC is unrelated to the PMT signals, it cannot be guaranteed that the real peak ($V_{p\_r}$) of the shaping circuit output signal is just located at one of the sampling points. The difference between $V_p$ and $V_{p\_r}$ is defined as peak error. With Eq. (1), we can calculate the RMS of the peak error of the $RC^2$ filter. In the circuit design, two sampling frequencies are considered, i.e. 40 MHz and 62.5 MHz. Fig. 6 shows the calculated result with different values of $\tau$, with the above two different sampling frequencies.

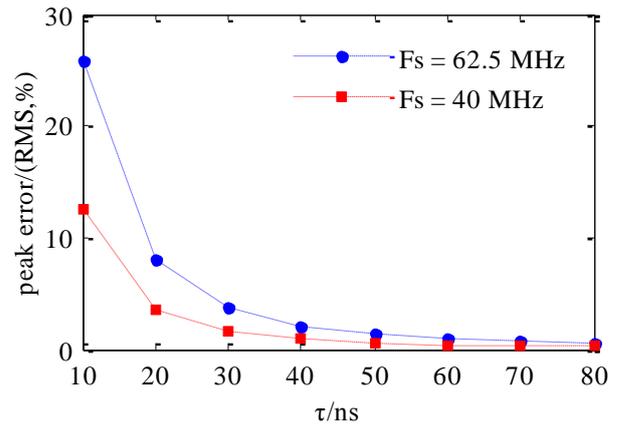

Fig. 6 The calculation results of the peak error of the $RC^2$ filter; $F_s$ is the sampling frequency of the ADC.

The Pspice simulation results are shown in Fig. 7. Both the calculation and simulation results indicate that larger value of $\tau$ would decrease the peak error, which contradict the requirement of $\tau$ as discussed in sub-section

3.1 Therefore, compromise has to be made to achieve a good measurement resolution. The calculation and simulation results also indicate that higher sampling rates would also decrease the peak error. In our design, a sampling frequency of 62.5 MHz is selected.

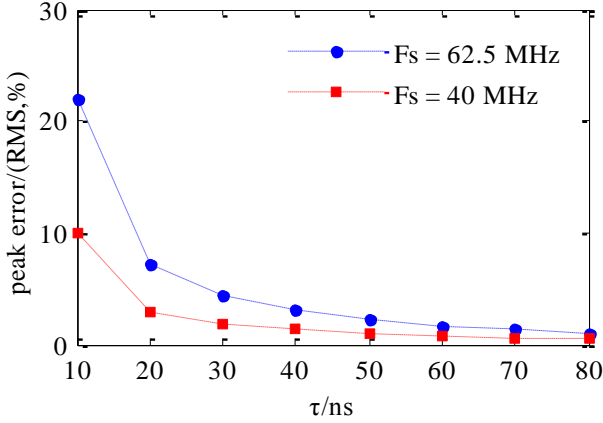

Fig. 7   Pspice simulation results of the peak error of the RC$^2$ filter; $F_s$ is the sampling frequency of the ADC.

### 3.3 Other aspects and optimizing the parameters

In the above sections, our analysis is based on the condition of impulse function input. Nevertheless, the rise time of real PMT signal would cause extra loss of the shaped signal height, which is termed Ballistic Deficit [15]. Increasing $\tau$ of the filter will decrease the Ballistic Deficit, however, meanwhile increasing the dead time of the charge measurement. Considering the pulse width of the input signal and the analysis result in sub-section 3.1 and 3.2, $\tau$ is selected to be 40 ns in our design. In this condition, according to the simulation results in Fig. 5 and Fig. 7, the SNR of the circuit is about 36 for the 1 P.E. input signal (equivalent to a charge resolution of 2.8%), while peak error is about 1%. Furthermore, the SNR of the circuit mainly influences the charge resolution of small signals while the peak error which is unrelated to the input signals mainly influence the charge resolution of large signals. Therefore, the simulation results indicate that the charge resolution is good enough for the application. Meanwhile, according to the Pspice transient waveform simulation, the charge measurement dead time is estimated to be about 400 ns, which is also acceptable.

Besides, the final charge measurement resolution would also be influenced by the performance of the ADC, which is usually characterized with the term of Effective Number Of Bits (ENOB). In this design, a 12 bit ADC (AD9222, 12 bits, from ADI Inc.) [16] is employed, and test results indicate that an ENOB of around 10 bits is achieved.

### 3.4 The charge measurement circuit of the AFE

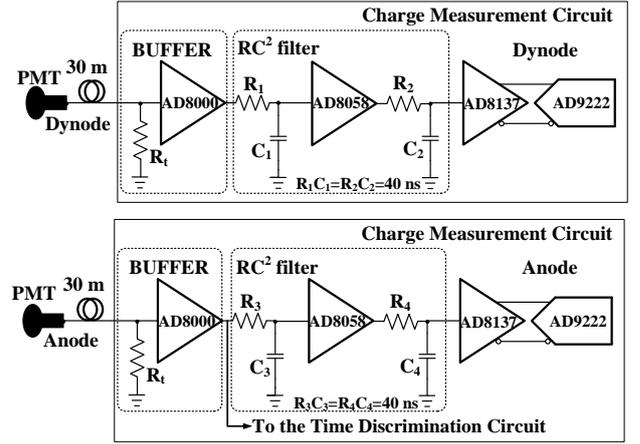

Fig. 8 The charge measurement circuit of the AFE for LHAASO WCDA.

According to the above analysis and simulation, the structure and primary component parameter are finally decided, as shown in Fig. 8. To simplify the circuit structure, the Anode and Dynode charge measurement circuits share a similar structure. A high-speed OPA (AD8000) [17] is employed as the pre-amplifier of Anode channel, the output of which is split into two paths: one is fed to the following charge measurement circuits, while the other is fed to the time discrimination circuit. The OPA used in each stage of the RC$^2$ filter is AD8058, which features a low noise and high bandwidth [18]. The output of RC$^2$ filter is imported to a differential amplifier (AD8137, from ADI Inc.) as the ADC driver. In addition, according to the above analysis, the $\tau$ is selected as 40 ns (400 $\Omega$+100 pF).

## 4   The design of the time discrimination circuit

The time measurement circuit is composed of the leading-edge discrimination circuit and FPGA TDC. In this paper, we focus on the design of the discrimination circuit which is crucial to the time resolution performance.

A current signal from the PMT is converted to a voltage signal by $R_t$ (50 Ω) in Fig. 9, and the voltage amplitude of the 1 P.E. signal is as low as 3 mV. To achieve a good time measurement resolution, the signal should be amplified with a gain high enough to guarantee the signal slew rate at the input of the discriminator. We designed an amplification circuit containing two stages of high-speed OPAs (AD8000) to obtain a high gain, as shown in Fig. 9. In order to avoid the ultra-deep saturation of the second OPA ($A_2$) with large input signals, we also placed a clamping circuit before it, which consists of $R_2$ and $D_2$ in Fig. 9.

The output of $A_2$ is AC coupled to filter out the baseline fluctuation, and then fed into two high speed discriminators (LMH7322, from TI Inc.) [19] with two different thresholds, as mentioned in Section-2. The low threshold is 1/4 PE, while the high threshold is set to be 5 PE, which can be controlled the FPGA through a 12-bit DAC (AD5628, from ADI Inc.) [20].

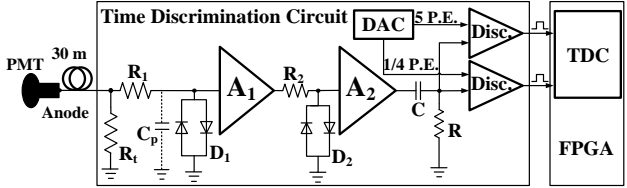

Fig. 9 Structure of the time measurement circuit.

As aforementioned, since the input signals ranges from 1 PE to 4000 PE, high precision impedance matching is required. The equivalent impedance at the input of the AFE shown in Fig. 9 can be expressed as:

$$Z_{eq} = R_t // (R_1 + \frac{1}{j\omega C_p} // R_{D1}) \quad , \qquad (6)$$

where $R_t$ is the matching resistor; $C_p$ is the front-end parasitic capacitance at the input of the $A_1$; $R_1$ is the resistance of the series resistor in the first clamping circuit, while $R_{D1}$ is the on-resistance of $D_1$.

With large input signals, the first clamping circuit works, which means a small value of $R_{D1}$. In this case, $R_1$ should be set large enough to guarantee the precision of impedance matching. Nevertheless, we also notice that $C_P$ and $R_1$ also form a RC filter which would reduce the slope at threshold-level crossing of the signal. In our design, $R_1$ is selected to be 500 Ω. Meanwhile, efforts should be made to deduce $C_p$. And this is the reason that we place one OPA (AD8000) as the pre-amplifier and then distribute the signal into two paths, as shown in Fig. 8. Of course, small package of the components at the input of the AFE are also considered.

Besides, we use the output of the low threshold discriminator as the start signal of the peak detection logic in FPGA. The ADC has a pipeline latency of 8 sampling periods [16]; we use this latency to calculate the baseline by averaging the first 8 ADC counts when the start signal is received. Then the logic seeks the maximum ADC count within 600 ns as the peak value. The charge measurement result is the value of difference between the peak value and baseline. This peak detection method can eliminate the baseline drift of the signal.

## 5 Initial laboratory tests of the AFE

A prototype of the AFE was designed and fabricated, as shown in Fig. 10, and a series of laboratory tests were conducted.

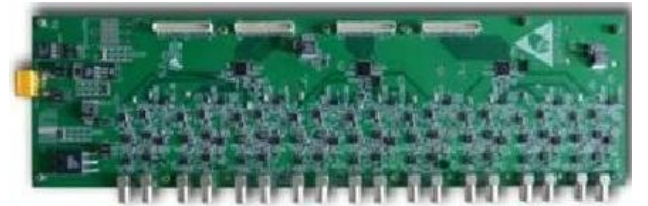

Fig. 10 A prototype of the LHAASO WCDA AFE

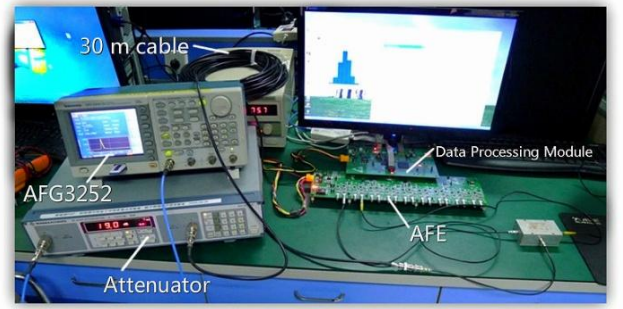

Fig. 11 The test bench in laboratory.

The system under test is shown in Fig. 11. We use an arbitrary signal source (AFG3252, from Tektronix Inc.) to generate the input signal, according to the waveforms of the PMT output signal using a high speed oscilloscope (DPO 7354C, from Tektronix Inc.). We tuned the signal amplitude through an external RF attenuator (from the ROHDE&SCHWARZ Inc.), and then the charge and time measurement performance can be tested with different input signal amplitudes. To approximate real application situation, the signal is imported to the AFE through a 30-meter cable (SYV50-2-41) from the attenuator. We use the Universal Serial Bus (USB) interface to readout the charge and time measurement results from the FPGA (placed in the Data Processing Module, as shown in Fig. 11), and analyze the data offline based on the MATLAB platform in the PC.

## 5.1 Charge Measurement test results

As mentioned above, by subtract the baseline calculation results from the peak detection result for each output digital pulse from the ADC, the charge measurement result can be obtained. By tuning the input signal amplitude, we can acquire the charge measurement performance with the input dynamic range.

Fig. 12 shows the charge measurement result (with unit of ADC count) with the different amplitudes. It can be observed that a dynamic range from 1 P.E. to 4000 P.E. is successfully achieved, and the regression coefficient of all channels is better than 0.9993. Fig. 13 shows the charge resolution test results, and the charge resolution is better than 10% @ 1 P.E. and 1% @ 4000 P.E., beyond the application requirement.

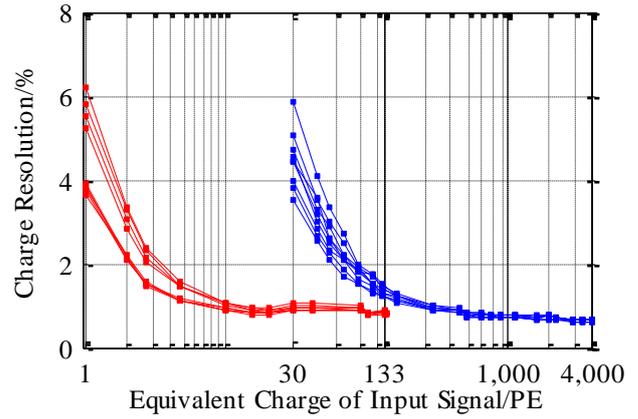

Fig. 13 charge resolution test results.

## 5.2 Time Measurement test results

Time resolution test is conducted based on the "cable delay test" method. The output signal of the RF attenuator is split and imported to two time measurement channels. Considering the situation that the time results of these two channels are not interrelated, the single-channel time resolution can be obtained by dividing by the RMS value of the time interval by $2^{1/2}$ [21].

We conducted the test in two steps. In the first step, we use the oscilloscope (Lecory 104MXi ) to obtain the time performance of the discrimination circuits in the AFE; in the second step, we use the FPGA TDC for time digitization to evaluate the time resolution of the whole system. The results are shown in Fig. 14 and Fig. 15.

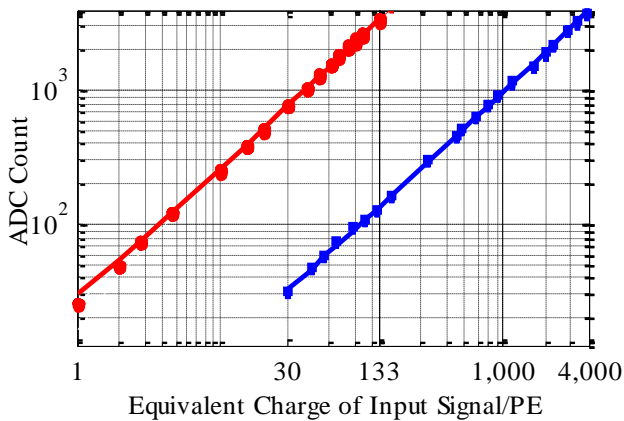

Fig. 12 the charge-ADC count conversion curve test results.

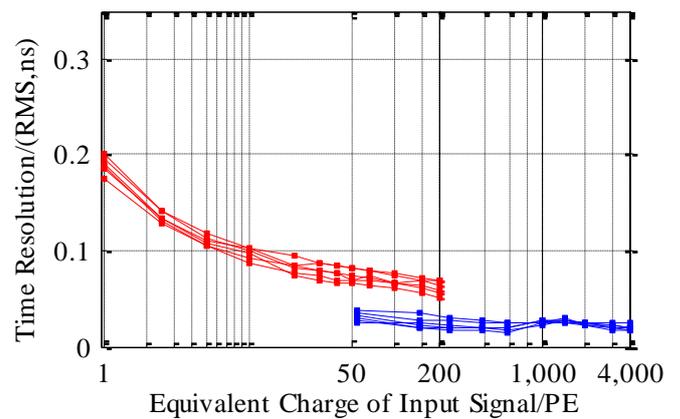

Fig. 14   the time resolution test results using oscilloscope.

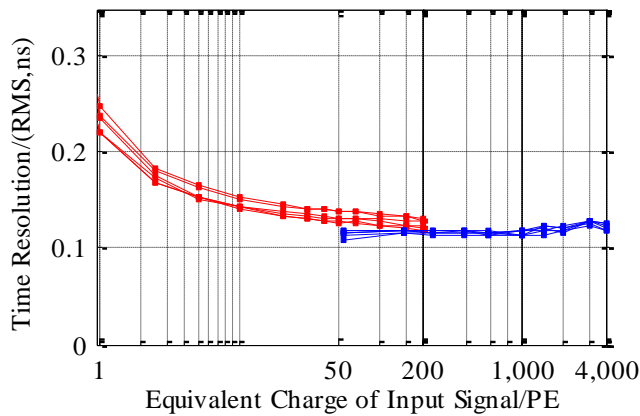

Fig. 15 the time resolution test results using FPGA TDC.

## 6    Summary

In this paper, we present the design of an AFE prototype for the WCDA in LHAASO. Structure and parameter optimization are included, with both analysis and simulations. Test were conducted to evaluate its performance, and the results indicate that a dynamic range from 1 P.E. to 4000 P.E. is achieved, and over the whole range, the charge resolution is better than 10 % @ 1 P.E. and 1% @ 4000 P.E., while the time resolution is better than 0.5 ns RMS, both beyond the application requirement.

*The authors would like to appreciate Dr. CAO Zhen, HE Hui-hai, YAO Zhi-guo, CHEN Ming-jun, LI Cheng and TANG Ze-bo for their kindly help. The authors thank all of the LHAASO WCDA collaborators who helped to make this work possible.*

As shown in Fig. 14 and Fig. 15, the time resolution is better than 0.5 ns RMS over the whole dynamic range, also beyond the requirement. The results in Fig. 15 are a little worse than those in Fig. 14, and the difference mainly comes from the quantification error of the FPGA TDC (bin size ~ 0.333 ns).